\documentclass[prl,showpacs,twocolumn]{revtex4}
\usepackage{stmaryrd}
\usepackage{times}
\usepackage{amsmath}
\usepackage{amsfonts}
\usepackage{amssymb}
\usepackage{graphicx}

\input{tcilatex}

\begin{document}

\title{Quantum Thermalization With Couplings }
\author{H. Dong$^{1}$ }
\author{S. Yang$^{1}$}
\author{X.F. Liu$^{2}$}
\author{C.P. Sun$^{1}$}
\email{suncp@itp.ac.cn}
\homepage{www.itp.ac.cn/~suncp}
\affiliation{$^{1}$Institute of Theoretical Physics,Chinese Academy of Science, Beijing
100080, China}
\affiliation{$^{2}$Department of Mathematics, Beijing University, Beijing 100871, China}

\begin{abstract}
We study the role of the system-bath coupling for the generalized
canonical thermalization [S. Popescu, et al., Nature Physics
2,754(2006) and S. Goldstein et al., Phys. Rev. Lett. 96,
050403(2006)] that reduces almost all the pure states of the
\textquotedblleft universe\textquotedblright\ [formed by a system $S
$ plus its surrounding heat bath $B$] to a canonical equilibrium
state of $S $. We present an exactly solvable, but universal model
for this kinematic thermalization with an explicit consideration
about the energy shell deformation due to the interaction between
$S$ and $B$. By calculating the state numbers of the
\textquotedblleft universe\textquotedblright\ and its subsystems $S$
and $B$ in various deformed energy shells, it is found that, for the
overwhelming majority of the \textquotedblleft
universe\textquotedblright\ states (they are entangled at least),
the diagonal canonical typicality remains robust with respect to
finite interactions between $S$ and $B$. Particularly, the kinematic
decoherence is utilized here to account for the vanishing of the
off-diagonal elements of the reduced density matrix of $S$. It is
pointed out that the non-vanishing off-diagonal elements due to the
finiteness of bath and the stronger system-bath interaction might
offer more novelties of the quantum thermalization.
\end{abstract}

\maketitle

Statistical mechanics is one of the most important and successful areas of
modern physics. However its foundation is still debatable and is actually
under debate. Most recently a mechanism for universal canonical
thermalization has been found in the following meaning: almost all the pure
states of the \textquotedblleft universe\textquotedblright\ consisting of
the\ considered system $S$ and  its \ surrounding heat bath $B$ can be
reduced into a generalized canonical state by tracing over the bath \cite%
{popescu}. Here, the allowed \textquotedblleft universe\textquotedblright\
states lie in a subspace defined by a general constraint $R$, which limits
the number of the states to be finite, yet very large. It is noticed that in
such generalized thermalization, the constraint $R$ is rather general and
need not be the energy shell of the \textquotedblleft
universe\textquotedblright , and generally speaking it doses not lead to the
usual canonical state in thermal equilibrium.

When the constraint $R$ is specialized as a total energy shell, the
generalized canonical state becomes a usual canonical thermal state. Such a
thermalization was described by the canonical typicality in ref. \cite%
{goldstein} associated with the so called overwhelming majority rule, based
on the law of large numbers, in counting the state numbers of the
constrained \textquotedblleft universe\textquotedblright\ and its
subsystems. Significant results concerning such canonical typicality have
been obtained by several authors for different purposes~\cite%
{bocchieri,lloyd1,tasaki,michel,breuer,gemmer}. Actually, to derive the
canonical distribution of $S$ from the microcanonical density matrix \cite%
{book2} or an entangled pure state \cite{goldstein} of the \textquotedblleft
universe\textquotedblright\ $U=S+B$, the interaction between $S$ and $B$
should be weak enough to allow a physical partition for the
\textquotedblleft universe\textquotedblright .\

In this paper, we will quantitatively consider the effect of system-bath
coupling on the above mentioned kinematics of canonical thermalization
leading to equilibrium canonical state. The present investigation only
concerns the usual canonical state. In the weak interaction limit, the
inverse temperature $\beta =\partial S\left( E\right) /\partial E$ emerges
from the thermodynamic entropy $S\left( E\right) =\ln \Omega \left( E,\delta
\right) $ where $\Omega \left( E,\delta \right) $ is the microstate number
of the bath in the energy shell $[E,E+\delta ]$. Since the interaction
between $S$ and $B$ deforms the geometry of the energy shell, we will
reexamine the validity of the temperature definition from the new
perspective offered by the generalized thermalization \cite%
{goldstein,popescu}. We try to understand how the off-diagonal elements of
the reduced density matrix of $S$ vanish due to the factorization structure,
in association with the random phase explanation \cite{sun1, sun2}. We find
that if the mode number of the bath is not large enough and the system-bath
coupling is strong enough, generally there exist nonzero off-diagonal
elements in the reduced density matrix. Notice that these nonzero
off-diagonal elements introduce quantum coherence into the usual thermal
equilibrium state and thus result in novel thermodynamic features.

We begin with a universal model: the system $S$ we consider is an $M$-level
system with the Hamiltonian $H_{S}=\sum \epsilon _{n}|n\rangle \langle n|$,
where $|n\rangle $ is the eigenstate with eigenvalue $\epsilon _{n}$, $%
n=1,2,..,M$; and the bath $B$ is modeled as a collection of $N$ harmonic
oscillators of frequencies $\omega _{j}\left( j=1,2,..N\right) $ with the
Hamiltonian $H_{B}=\sum_{j}\omega _{j}a_{j}^{\dagger }a_{j}$. This model can
be regarded as a universal approach, because in the weak coupling limit, any
heat bath could be universally modeled as a collection of harmonic
oscillators with the linear couplings to the surrounded system according to
the proofs in ref. \cite{legget}. For this reason, the interaction $H_{I}$
between $B$ and $S$ should be modeled to be linear with respect to the bath
variables $a_{j}^{\dagger }$ and $a_{j}$. So we assume a simplest
system-bath coupling
\begin{equation}
H_{I}=\sum_{j,n}\lambda _{n}|n\rangle \langle n|(g_{j}a_{j}^{\dagger }+\text{%
H.c.)},
\end{equation}%
where $\lambda _{n}$ are real numbers. Note that $H_{I}$ is of
non-demolition since $[H_{S},H_{I}]=0$. Thus the interaction only causes the
dephasing of $S$, and the energy dissipation of $S$ will not appear \cite%
{sun2,gao}.

Obviously, the eigenvalues of the \textquotedblleft
universe\textquotedblright\ formed by $B$ and $S$ are
\begin{equation}
E\left( n,\{n_{j}\}\right) =\epsilon _{n}\left( \kappa \right)
+\sum_{j=1}^{N}n_{j}\omega _{j},  \label{energy}
\end{equation}%
corresponding to the eigenstates $\left\vert n,\{n_{j}\}\right\rangle
=\left\vert n\right\rangle \otimes \prod_{j=1}^{N}\left\vert n_{j}\left(
n\right) \right\rangle $, where $\left\vert n_{j}\left( n\right)
\right\rangle =D\left( \alpha _{jn}\right) \left\vert n_{j}\right\rangle $
is defined in terms of the Fock states $\left\vert n_{j}\right\rangle $ of $%
B $ and the coherent-state-generating operator $D\left( \alpha _{jn}\right)
=\exp (\alpha _{jn}a_{j}^{\dagger }-h.c)$ with the displacement parameters $%
\alpha _{jn}=-\lambda _{n}g_{j}/\left( 2\omega _{j}\right) $; $\epsilon
_{n}\left( \kappa \right) =\epsilon _{n}-\kappa \lambda _{n}^{2}$.
Especially, the parameter $\kappa =\Sigma _{j}\left\vert g_{j}\right\vert
^{2}/\left( 4\omega _{j}\right) $ reflects the role of the interaction
between $S$ and $B $.

\begin{figure}[tbp]
\includegraphics[width=8 cm, clip]{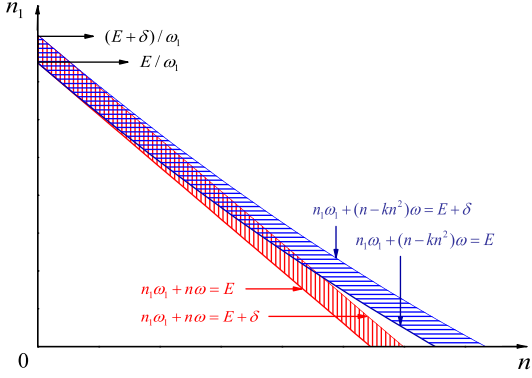}
\caption{\textit{(Color Online) Geometrical deformation of the energy shell.
Here we show explicitly the deformation, when $S$ is a harmonic oscillator.
(a) Without the interaction, the energy shell is the red area; (b) The
interaction deforms the red area into the blue area.}}
\label{Concavity}
\end{figure}

Note that the system-bath coupling deforms the energy shell of thickness $%
\delta $ defined by the total constraint $E\leq E\left( n,\{n_{j}\}\right)
\leq E+\delta $. This energy shell determines a subset of the
\textquotedblleft universe" states. For convenience, we denote by $V\left(
E,\delta \right) $ the subspace spanned by this subset. To have a clear
picture of this deformation of energy shell, let us consider the following
simple example: the system $S$ is a harmonic oscillator with frequency $%
\omega $ and $\lambda _{n}=n.$ In this case, the renormalized energy of $S$
is $E_{s}\left( n\right) =n\omega -\kappa n^{2}$. In Fig. \ref{Concavity},
we illustrate the deformation of the energy shell when $N=1$. If there were
no interaction, the above constraint would produce the red area. The
interaction introduces the nonlinear term $\kappa n^{2}$ to deform it into
the blue area.

Generally, in order to derive the canonical distribution of $S$ from an
entangled pure state of the \textquotedblleft universe\textquotedblright ,
we need to calculate the dimension $\Omega _{N+1}\left( E,\delta ,\kappa
\right) $ of$\ V\left( E,\delta \right) $ and the dimension $\Omega
_{N}\left( E-\epsilon _{n},\delta ,\kappa \right) $ of the subspace $%
V^{B}\left( n,\kappa \right) \ $spanned by the states with the constraint%
\begin{equation}
E-\epsilon _{n}\left( \kappa \right) \leq \sum_{j}n_{j}\omega _{j}\leq
E+\delta -\epsilon _{n}\left( \kappa \right) \text{ .}  \label{constrain2}
\end{equation}

Let us start with a pure entanglement state $\left\vert \psi \right\rangle
=\sum^{\prime }C\left( n,\{n_{j}\}\right) \left\vert
n,\{n_{j}\}\right\rangle $ of the constrained \textquotedblleft
universe\textquotedblright , where $\sum^{\prime }$ denotes the summation
under the total constraint. By tracing over the variables of $B$, the
reduced density matrix $\rho _{S}=Tr_{B}\left( \left\vert \psi \right\rangle
\left\langle \psi \right\vert \right) $ of $S$ is obtained as
\begin{equation}
\rho _{S}=\sum_{n}P_{n}\left\vert n\right\rangle \left\langle n\right\vert
+\sum_{n\neq m}F_{nm}\left\vert n\right\rangle \left\langle m\right\vert ,
\end{equation}%
where the diagonal elements are $P_{n}=P_{n}\left( E,\kappa \right)
=\sum^{\prime \prime }\left\vert C\left( n,n_{j}\right) \right\vert ^{2}$
and $\sum^{\prime \prime }$ stands for the summation under the constraint (%
\ref{constrain2}). We represent this summation by the summation over the
constrained indices $[n_{j}]_{E,n}$,\ then the off-diagonal elements can be
written as
\begin{equation}
F_{nm}=\sum_{[m_{j}]_{E,m}}\sum_{[n_{j}]_{E,n}}C\left( n,n_{j}\right)
C^{\ast }\left( m,m_{j}\right) D_{m\left( m_{j}\right) }^{n\left(
n_{j}\right) }.  \label{dec}
\end{equation}%
where
\begin{equation}
D_{m\left( m_{j}\right) }^{n\left( n_{j}\right)
}=\prod_{j=1}^{N}d_{m_{j}\left( m\right) }^{n_{j}\left( n\right)
}=\prod_{j=1}^{N}\left\langle m_{j}\left( m\right) \right. \left\vert
n_{j}\left( n\right) \right\rangle
\end{equation}%
are decoherence factors with a factorized structure and for $m\neq
n,m_{j}\geq n_{j},$ each factor
\begin{equation}
d_{m_{j}\left( m\right) }^{n_{j}\left( n\right) }=\Delta _{\alpha }^{\left(
m_{j}-n_{j}\right) }e^{-\Delta _{\alpha }^{2}/2}L_{n_{j}}^{\left(
m_{j}-n_{j}\right) }\left( \Delta _{\alpha }^{2}\right) \sqrt{\frac{n_{j}!}{%
m_{j}!}}
\end{equation}%
is expressed in terms of the associated Laguerre polynomials$\
L_{n}^{m}\left( x\right) $ with the variable $\Delta _{\alpha
}=-g_{j}/\left( 2\omega _{j}\right) \left( n-m\right) $.

In order to show the generalized quantum thermalization, we need to study
how the diagonal elements of $\rho _{S}$\ approach the Gibbs distribution $%
P_{n}\varpropto p_{n}^{G}=\exp (-\beta \epsilon _{n})$, while the
off-diagonal elements $F_{nm}$ vanish as a quantum decoherence effect. To
this end, we use the basic assumptions for statistical mechanics that $%
\delta \ll \omega $, but $\delta \gg \omega _{j}$ and $N\rightarrow \infty $.

First, we consider the diagonal elements. Let us establish the formula%
\begin{equation}
P_{n}\left( E,\kappa \right) =\frac{\Omega _{N}\left( E-\epsilon _{n},\delta
,\kappa \right) }{\Omega _{N+1}\left( E,\delta ,\kappa \right) }.  \label{PG}
\end{equation}%
in some sense under the reasonable assumption that $\left\vert C\left(
n,\{n_{j}\}\right) \right\vert ^{2}$ are random variables with an identical
distribution. In fact, if this condition is satisfied, then these random
variables have the same mathematical expectation $\left( \Omega _{N+1}\left(
E,\delta ,\kappa \right) \right) ^{-1}$ due to the restriction $%
1=\sum^{\prime }\left\vert C\left( n,\{n_{j}\}\right) \right\vert ^{2}$. It
then follows from the law of large numbers that $\sum^{\prime \prime
}\left\vert C\left( n,\{n_{j}\}\right) \right\vert ^{2}$ approaches $\Omega
_{N}\left( E-\epsilon _{n},\delta ,\kappa \right) \left( \Omega _{N+1}\left(
E,\delta ,\kappa \right) \right) ^{-1}$ in probability. Namely, the above
formula is valid with high probability. Here we remark that with a
straightforward calculation (e.g., in Ref. \cite{goldstein}), this formula
can be obtained from the microcanonical state $\rho _{MC}=\sum^{\prime
}[1/\Omega _{N+1}\left( E,\delta ,\kappa \right) ]\left\vert
n,\{n_{j}\}\right\rangle \left\langle n,\{n_{j}\}\right\vert $ of the
\textquotedblleft universe\textquotedblright .

We are now in a position to consider how $P_{n}\left( E,\kappa \right) $
leads to the canonical distribution. According to Refs. \cite%
{goldstein,popescu}, when there is no interaction, $P_{n}\left( E,\kappa
=0\right) $ does give rise to the canonical distribution for almost all the
pure states $\left\vert \psi \right\rangle $ of the constrained
\textquotedblleft universe\textquotedblright . But when there exists an
interaction between $S$ and $B$ , is it still the case? The answer seems to
be positive when the interaction is weak. We will attack this problem by
theoretical analysis and numerical simulation.

If we can show that $P_{n}\left( E,\kappa \right) $ possesses a
\textquotedblleft conformal invariance\textquotedblright\ with respect to
the geometrical deformation of the energy shell caused by the interaction,
i.e., $P_{n}\left( E,\kappa \right) \sim P_{n}\left( E,\kappa =0\right) $,
then the problem is solved. We try to justify this \textquotedblleft
conformal invariance\textquotedblright\ for weak interaction ($\kappa \neq 0$%
). It follows from the direct sum decomposition $V\left( E,\delta \right)
=\sum \oplus V^{B}\left( n,\kappa \right) $ of the Hilbert space $V\left(
E,\delta \right) $ that the dimension of $V\left( E,\delta \right) $ can be
written as $\Omega _{N+1}\left( E,\delta ,\kappa \right)
=\sum_{n=1}^{M}\Omega _{N}(n)$ . Here, $\Omega _{N}(n)$ stands for the
number of states in the area defined by Eq. (\ref{constrain2}), and $M$ is
an upper bound of the summation range, which is determined by the
positiveness of both the bath energy and the system energy. By
straightforward calculation we obtain
\begin{equation}
\Omega _{N}(n)\simeq \frac{\left( E-\epsilon _{n}\left( \kappa \right)
\right) ^{N-1}\delta }{\left( N-1\right) !\prod_{j=1}^{N}\omega _{j}^{2}}
\end{equation}%
for very small $\delta $. Thus the diagonal elements of $\rho _{S}$ take the
form
\begin{equation}
P_{n}=\frac{[E-\epsilon _{n}\left( \kappa \right) ]^{N-1}}{%
\sum_{n=1}^{M}[E-\epsilon _{n}\left( \kappa \right) ]^{N-1}}.  \label{pn}
\end{equation}%
with a finite $M$ and a large\textbf{\ }$N$. Since the eigen-energy $%
\epsilon _{n}$ of the system is much smaller than the total energy of the
energy shell, we have $[E-\epsilon _{n}\left( \kappa \right) ]^{N-1}\approx
\exp [(N-1)\ln (E+\kappa \lambda _{n}^{2})]\exp (-\beta _{n}\epsilon _{n})$
where $\beta _{n}$ is the quasi-temperature defined as

\begin{equation}
\beta _{n}=\frac{\partial S\left( E\right) }{\partial E}|_{E\rightarrow
E+\kappa \lambda _{n}^{2}}=\frac{N-1}{E+\kappa \lambda _{n}^{2}}.
\end{equation}%
Usually the energy correction $\kappa \lambda _{n}^{2}=\lambda
_{n}^{2}\Sigma _{j}\left\vert g_{j}\right\vert ^{2}/\left( 4\omega
_{j}\right) $ is much smaller than the shell energy $E,$ thus the
quasi-temperature becomes independent of $n$: $\beta _{n}\approx N/E$.
Therefore, $P_{n}$ is \textquotedblleft conformably invariant" with respect
to $\kappa $, and for this reason we can expect the Gibbs distribution $%
P_{n}=\exp (-\beta _{n}\epsilon _{n})/[\sum_{n=1}^{M}\exp (-\beta
_{n}\epsilon _{n})].$

\begin{figure}[h]
\includegraphics[bb=17 17 275 215, width=8 cm, clip]{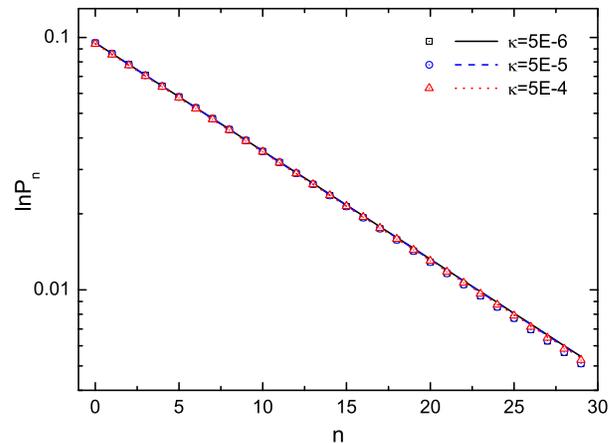}
\caption{\textit{(Color Online) Plot of $lnP_{n}$ as a function of the
number $n$ of the system, when $N=50$ with different $\protect\kappa$: $%
5\times10^{-6}$($\boxempty$ ), $5\times 10^{-5}$($\bigcirc$)\ and $%
5\times10^{-4}$($\triangle$). The fitting of the Gibbs distribution is shown
by the corresponding lines.}}
\label{fig:pn}
\end{figure}

It should be pointed out that the above argument is only heuristic since
neither the sum $\sum \Omega _{N}(n)$\ nor the distribution $P_{n}$\ is
calculated analytically. Thus we will resort to numerical simulation to
support our expectation. Let us assume the system is a harmonic oscillator.
In this case, $\lambda _{n}=n$ and the quasi-temperature of the system is
exactly $\beta _{n}=\left( N-1\right) /\left( E-\kappa n^{2}\right) $. Due
to the weakness of interaction, $\beta _{n}$ is approximately equal to $N/E$
when $n$ is small. We choose $E=0.5$ and $\omega =10^{-3}$ and display in
Fig.\ref{fig:pn} the relationship between the distribution $P_{n}$ and $n$
for different $\kappa $ when $N=50.$ It is clearly shown that $P_{n}$ indeed
decays exponentially as Gibbs distribution as $n$ increases. Tab. \ref%
{table:temp} gives the inverse temperature of the system for different
coupling parameters $\kappa $. Here, the theoretical inverse temperature is $%
\beta =98$. These numerical results demonstrate that the state obtained by
tracing over the bath is just the Gibbs canonical state under some
conditions.
\begin{table}[h]
\caption{Inverse Temperature vs Interaction }
\label{table:temp}\tabcolsep 5mm
\par
\begin{center}
\begin{tabular}{llll}
\hline
$\kappa$ & $5\times10^{-6}$ & $5\times10^{-5}$ & $5\times10^{-4}$ \\ \hline
$\beta$ & 98.94 & 98.85 & 98.69 \\ \hline
\end{tabular}%
\end{center}
\end{table}

Now we consider the bath-induced decoherence effect, which causes the
off-diagonal elements of $\rho _{S}$\ to approach zero. When the system-bath
couplings can be ignored in comparison with the energy level spacing and the
total energy of the bath, any two subspaces defined by the constraint (\ref%
{constrain2}) can not overlap each other and then the decoherence factor
vanishes. In present situation, the system-bath coupling will weaken this
decoherence for thermalization since the positive terms $\lambda
_{n+1}^{2}-\lambda _{n-1}^{2}$ reduce the effective system's energy spacings
\begin{equation}
\epsilon _{n+1}(\kappa )-\epsilon _{n}(\kappa )=\epsilon _{n+1}-\epsilon
_{n}-\kappa (\lambda _{n+1}^{2}-\lambda _{n-1}^{2})
\end{equation}%
to make them comparable to the thickness $\delta $ of the energy shell. Then
two subspaces $V^{B}\left( n,\kappa \right) $ and $V^{B}\left( m,\kappa
\right) $ of $B$ can overlap each other and the off-diagonal elements $%
F_{nm} $ will not vanish. However, notice that the norm of each component $%
d_{m_{j}\left( m\right) }^{n_{j}\left( n\right) }$ in the decoherence factor
is less than unity. Thus the decoherence factor $D$ may still vanish in the
thermodynamic limit $N\rightarrow \infty .$ So we can say that the
factorized structure of $D$ enhances the decoherence \cite{sun1, sun2}, and
to some extent compensates the negative effect of interaction in
thermalization.

Next, we wish to point out that in the mesoscopic case, that is to say, $N$
is not large enough, novel effects may arise. In this case, if the
system-bath coupling is strong, there will exist finite off-diagonal
elements $F_{nm}$ in the reduced density matrix. This means quantum
coherence is introduced into the usual thermally equilibrium state. Such a
state is called a quasi-thermal state.

For a two level system with single energy spacing $\Delta $, the
quasi-thermal state can be described by the reduced density matrix
\begin{equation}
\rho _{S}=\left[
\begin{array}{cc}
p_{+} & F \\
F^{\ast } & p_{-}%
\end{array}%
\right]
\end{equation}%
The diagonal elements $p_{\pm }=1/(1+\exp (\pm \beta \Delta ))$ approach the
standard Gibbs distributions while the off-diagonal elements are
non-vanishing: $F=F_{12}.$ We can diagonalize the above reduced density
matrix to obtain the two effective probabilities $P_{\pm }(F)$ exactly. For
small $F$ , $P_{\pm }(F)$ can be approximated as
\begin{equation}
P_{\pm }(F)\approx p_{\pm }\mp \coth (\frac{\beta \Delta }{2})\left\vert
F\right\vert ^{2}
\end{equation}%
Then the von Neumann entropy $S_{VN}=-\sum_{\alpha =\pm }P_{\alpha }(F)\ln
P_{\alpha }(F)$ is approximated as
\begin{equation}
S_{VN}\simeq S\left( E\right) -\beta \Delta \left\vert F\right\vert
^{2}\coth (\frac{\beta \Delta }{2})
\end{equation}%
It is observed that due to the system-bath interaction the von Neumann
entropy explicitly deviates from the thermodynamic entropy

\begin{equation*}
S\left( E\right) \simeq \frac{\beta \Delta }{e^{\beta \Delta }+1}+\ln
(e^{-\beta \Delta }+1),
\end{equation*}%
which is by definition the entropy of the Gibbs equilibrium state $\rho
_{G}=diag(p_{+},p_{-})$. Therefore, generally von Neumann entropy does not
relate to the meaningful usual notion of temperature. In fact, there exists
no good physical notion of temperature for a general
non-thermal-equilibrium. But the erasure of quantum information by
thermalization indicates where temperature enters in this matter.

However, only for two-level system or system with homogeneous energy level
spacing can we define an effective temperature $T_{\mathrm{eff}}=1/\beta _{%
\mathrm{eff}}$ \cite{quan} by the ratio $r(t)=P_{+}(F)/P_{-}(F)=\exp (-\beta
_{\mathrm{eff}}\Delta )$ and the level spacing $\Delta $. For such a system
interacting with a finite heat bath, even in a non equilibrium state, we can
imagine that it is in a virtual equilibrium state with the effective inverse
temperature
\begin{equation}
\beta _{\mathrm{eff}}\simeq \beta +\frac{4\left\vert F\right\vert ^{2}}{%
\Delta }\cosh ^{2}(\frac{\beta \Delta }{2})\coth (\frac{\beta \Delta }{2})
\end{equation}%
This effective temperature in the quasi-thermal state is higher than the
usual equilibrium temperature. Such kind of quasi-thermal state with a bit
of quantum coherence can demonstrate various exotic natures in
thermodynamical processes. Scully et al \cite{scully} have proposed a
quantum Carnot engine in which the bath atoms are given some quantum
coherence, which can increase the effective temperature of the radiation
field. In this case, though the second law of thermodynamics is not
violated, the quantum Carnot engine possesses some features that are not
possible in a classical case.

In summary, the quantum kinetic thermalization of the system is explored,
based on an exactly solved and general model with weak interaction between $%
S $ and $B$ and the effect of interaction is demonstrated as the deformation
of the energy shell; Based on the model, we realized the thermalization by
tracing over the variable of the bath $B$. Here decoherence is used to
account for the disappearance of the off-diagonal elements of the reduced
density matrix of the system in contact with a bath of infinitely large
particle number in the weak coupling limit. Moreover, the novel
thermodynamic effects are emphasized that can result from the non-vanishing
off-diagonal elements of the reduced density matrix when the bath is
mesoscopic.

This work is supported by the NSFC with grant Nos. 90203018, 10474104 and
60433050, and NFRPC with Nos. 2006CB921206 and 2005CB724508.

\end{document}